\documentclass[smallextended]{svjour3}       
\smartqed  
\usepackage{graphicx}
\usepackage{amsmath}
\usepackage{amsfonts}
\usepackage{amssymb}
\usepackage{graphicx}
\usepackage{cite}
\usepackage{color}

	\newcommand{\ket}[1]{\left| #1 \right\rangle}
	\newcommand{\bra}[1]{\left\langle #1 \right|}

\begin{document}

\title{The effect of classical driving field on the spectrum of a qubit and entanglement swapping inside dissipative cavities}

\titlerunning{The effect of classical driving field on the spectrum ...}        

\author{Ali Mortezapour  \and Alireza Nourmandipour  \and Hossein Gholipour 
}


\institute{Ali Mortezapour \at
              Department of Physics, University of Guilan, P. O. 41335–1914, Rasht, Iran \\
              \email{mortezapour@guilan.ac.ir}
           \and 
           Alireza Nourmandipour \at
           Department of Physics, Sirjan University of Technology, Sirjan, Iran \\
           \email{anourmandip@sirjantech.ac.ir}
            \and  
            Hossein Gholipour \at
              Department of Physics, University of Guilan, P. O. 41335–1914, Rasht, Iran \\
}

\date{Received: date / Accepted: date}

\maketitle

\begin{abstract}

In this paper, we study the effect of classical driving field on the spontaneous emission spectrum of a qubit embedded in a dissipative cavity. Furthermore, we monitor the entanglement dynamics of the driven qubit with its radiative decay under the action of the classical field. Afterwards, we carry out an investigation on the possibility of entanglement swapping between two such distinct driven qubits. The swapping will be feasible with the aid of a Bell state measurement performing on the photons leaving the cavities. It is demonstrated that the classical driving field has a beneficial effect on the prolonging of the swapped entanglement.

\end{abstract}

\section{Introduction}
\label{intro}
The quantum phenomenon of entanglement is an intrinsic non-separability feature of states which arises due to the superposition principle in multipartite systems. Hence, entanglement can leak out valuable information about composite systems. On the other hand, entanglement undertakes a leading role in quantum information processing. Indeed, entanglement is recognized as a fundamental resource for quantum information tasks such as quantum teleportation \cite{Ali1}, quantum computation \cite{Ali2}, quantum cryptography \cite{Ali3,Ali4} and superdense coding \cite{Ali5}. In this regard, various schemes have been proposed for creating and detecting a desirable entangled state in quantum systems. The proposed schemes mainly include atoms (real or artificial) in optical cavities \cite{Ali6}, quantum dots \cite{Ali7}, trapped ions \cite{Ali8,Ali9,Ali10,Ali11} and superconducting quantum interference devices \cite{Ali12,Ali13}. 

All of the entangling schemes rely on direct or indirect interactions between systems. Atom-photon interaction known as a direct method for creating entanglement. On the contrary to the direct interaction, it is possible to create entanglement between subsystems distributed over long distances without any common past, the phenomenon called entanglement swapping \cite{Alireza1}. This notion was originally proposed to swap entanglement between a pair of particles \cite{Alireza1}. After that, the entanglement swapping was generalized for continuous variable systems \cite{Alireza2}. In \cite{Alireza3}, the authors have experimentally demonstrated the unconditional entanglement swapping for continuous variables. Loss-resistant state teleportation and entanglement swapping using a quantum-dot spin in an optical microcavity has also been studied \cite{Alireza4}. Entanglement swapping in two independent Jaynes Cummings model  has been discussed in \cite{Alireza5}. The influence of detuning  and Kerr medium on the entanglement swapping has been investigated in \cite{Alireza6}. Recently, it has been shown that the entanglement can be swapped between dissipative systems \cite{Alireza7}.

It is well-known that entanglement in any realistic systems becomes fragile due to the inevitable interaction between the system and its surrounding environment \cite{Ali20,Ali21,Ali22}. Meanwhile, a successful quantum operation requires quantum systems with long-lasting entanglement. Along this route, some strategies have been put forward to control the time-evolution of entanglement and shield it from decay \cite{Ali23,Ali24,Ali25,Alireza8,Alireza9,Ali26,Ali27,Ali28,Ali29,Ali30,Ali31,Ali32,Alireza10,Alireza11,Alireza12,Ali33,Ali34,Ali35,Ali36,Ali37,Ali38,Ali39,Ali40,Ali41,Ali42,Ali43,Ali44,Ali45}. However, controlled manipulation of individual quantum systems is one of the most remarkable recent achievements of experimental science. Nowadays, individual atoms and photons can be guided through complex coherent evolutions with dominant control. One of the most effective control, which can be realized in both cavity-QED and circuit-QED setups, is classical control.
In this work, we focus on a model of a qubit interacting with a zero-temperature structured reservoir and driven by an external classical control field. It was shown that the entanglement between two such qubits, which are embedded in two independent cavities, can be preserved for a long time \cite{Ali46}. Moreover, non-Markovianity \cite{Ali47,Ali48,Ali49}, quantum Fisher information \cite{Ali50,Ali51}, quantum speedup \cite{Ali48}, coherence dynamics \cite{Ali52} and quantumness \cite{Ali49} of such a model have been studied.

 Here, we aim to study the spontaneous emission spectrum of the qubit as well as the entanglement dynamic between the qubit and its radiative decay under the classical control. 
We also go further and study the possibility of entanglement swapping between such systems. To end this, we consider two similar systems each composed of a single qubit interacting with its own surrounding environment under the classical control. After having obtained the wave function of each system, we perform a Bell state measurement on the fields leaving the cavities which leaves qubits in an entangled state. We use the concurrence \cite{Alireza13} to quantify the amount of swapped entanglement. We then take a statistical average over the initial states of the qubits to establish an input-independent dynamics of entanglement, the concept of entangling power \cite{Alireza14}.

The paper is organized as follows. In Sec. \ref{ModSol} we introduce the model and give the explicit expression of the evolved reduced density matrix. The results regarding spontaneous emission spectrum are presented in Sec. \ref{Spe}. In Sec. \ref{atmphy}, using von-Neumann entropy, we discuss the time evolution of entanglement between the driven qubit and its radiative decay. 
In Sec. \ref{EntSwa}, entanglement swapping between such systems is discussed. 
Finally, in Sec. \ref{Con} we summarize our conclusions.

\section{Model and its Solution}
\label{ModSol}

\begin{figure}[h!]
   \centering
\includegraphics[width=0.6\textwidth]{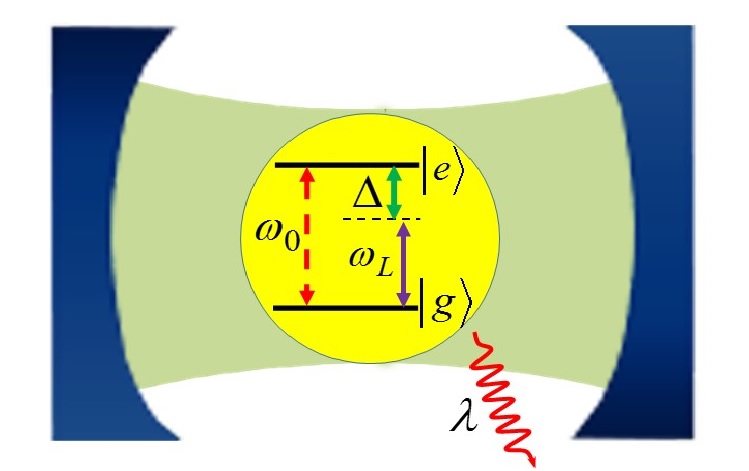}
   \caption{\label{Fig1} Schematic illustration of a setup in which a qubit is driving by a classical field inside a leaky cavity.}
  \end{figure}

Consider a qubit (two-level system) with transition frequency $\omega_0$ which is driven by an external classical field and embedded in a zero-temperature reservoir formed by the quantized modes of a high-Q cavity as depicted in Fig. \ref{Fig1}. Under the dipole and rotating-wave approximations, the associated Hamiltonian of the system can be written as ($\hbar=1$):
\begin{equation}
\label{Eq1}
\begin{aligned}
\hat{H}&=\frac{\omega_0}{2} \hat{\sigma}_z+\sum_{k}\omega_{k} {\hat{a}_{k}}^{\dagger}\hat{a}_{k}\\ &+\sum_{k}g_{k}\hat{a}_{k}\hat{\sigma}_{+}+\Omega e^{-i\omega _Lt}\hat{\sigma}_{+}+\text{H.c.}
\end{aligned}
\end{equation}
where ${{\hat{\sigma }}_{z}}=\left| e \right\rangle \left\langle  e \right|-\left| g \right\rangle \left\langle  g \right|$ is the Pauli matrix, ${{\omega }_{L}}$ and ${{\omega }_{k}}$  represent the frequencies of  the classical driving field and the cavity quantized modes, respectively. ${{\hat{\sigma }}_{+}}=\left| e \right\rangle \left\langle  g \right|$ (${{\hat{\sigma }}_{-}}=\left| g \right\rangle \left\langle  e \right|$) denotes the qubit raising (lowering) operator, while ${{\hat{a}}_{k}}$ ($\hat{a}_{k}^{\dagger}$) are the annihilation (creation) operators of the cavity $k$th mode. In addition, $\Omega $ and ${{g}_{k}}$ represent the coupling strength of the interactions of the qubit with the classical driving field and the cavity modes, respectively. We assume that $\Omega $ is a real number and to be small compared to the atomic and laser frequencies ($\Omega <<{{\omega }_{0}},{{\omega }_{L}}$).
Making use of the unitary transformation $U={{e}^{-i{{\omega }_{L}}{{{\hat{\sigma }}}_{z}}t/2}}$, the Hamiltonian of the system in the rotating reference frame can be written as:
\begin{equation}
\label{Eq2}
\begin{aligned}
\hat{H}_{\text{eff}}&=\hat{H}_{\text{I}}+\hat{H}_{\text{II}}, \\
\hat{H}_{\text{I}}&=\frac{\Delta}{2}\hat{\sigma}_{z}+\Omega\hat{\sigma}_{x}, \\
\hat{H}_{\text{II}}&=\sum_{k}\omega_k\hat{a}_k^{\dagger}\hat{a}_k+ \sum_{k}\left\lbrace g_{k}\hat{a}_{k}e^{+i\omega_Lt}\hat{\sigma}_{+}+\text{H.c.}\right\rbrace,
\end{aligned}
\end{equation}
in which $\Delta ={{\omega }_{0}}-{{\omega }_{L}}$ is the detuning between the qubit and the classical driving field. By introducing the dressed states

\begin{equation}
\begin{aligned}
\ket{E}&=\sin\frac{\eta}{2}\ket{g}+\cos\frac{\eta}{2}\ket{e},\\
\ket{G}&=\cos\frac{\eta}{2}\ket{g}-\sin\frac{\eta}{2}\ket{e},
\end{aligned}
\end{equation}
which are the eigenstates of ${{\hat{H}}_{\text{I}}}$, the effective Hamiltonian can be re-written as
\begin{equation}
\begin{aligned}
\hat{H}_{\text{eff}}&=\frac{\omega_D}{2}\hat{\chi}_{z}+\sum_{k}\omega _{k}\hat{a}_{k}^{\dagger}\hat{a}_{k}\\
&+{{\cos }^{2}}(\eta /2)\sum_{k}\left\lbrace {g}_{k}\hat{a}_{k}\hat{\chi}_{+}e^{+i{{\omega }_{L}}t}+\text{H.c.}\right\rbrace.
\end{aligned}
\end{equation}
Where ${{\hat{\chi }}_{z}}=\left| E \right\rangle \left\langle  E \right|-\left| G \right\rangle \left\langle  G \right|$ denotes the new Pauli matrix and ${{\omega }_{D}}=\sqrt{{{\Delta }^{2}}+4{{\Omega }^{2}}}$ being the dressed qubit frequency. Moreover, $\eta =\operatorname{Arctan}[2\Omega /\Delta ]$ and ${{\hat{\chi }}_{+}}=\left| E \right\rangle \left\langle  G \right|$ ($({{\hat{\chi }}_{-}}=\left| G \right\rangle \left\langle  E \right|\ $) represents the new lowering (raising) operator. It is noteworthy that 
we have neglected the terms of non-conservation of energy by using the usual rotating-wave approximation\cite{Ali46,Ali52}. 

We assume the overall system to be initially in a product state with the qubit in a coherent superposition of its states ($\cos(\theta/2) \ket{E}+\sin(\theta/2) e^{i\phi}\ket{G} $) and the reservoir modes in the vacuum state $\ket{\boldsymbol{0}}_{R}$, i.e.,
\begin{equation}
\ket{\psi(0)}=\left( \cos(\theta/2) \ket{E}+\sin(\theta/2) e^{i\phi}\ket{G}\right)\ket{\boldsymbol{0}}_{R}.
\label{eq:initialstate}
\end{equation}
Hence, the state vector of the system at any time $t$ can be written as 
\begin{equation}
\begin{aligned}
\ket{\psi(t)}&=\cos(\theta/2){\cal E}(t)\ket{E}\ket{\boldsymbol{0}}_{R}+\sin(\theta/2) e^{i\phi}\ket{G}\ket{\boldsymbol{0}}_{R} \\
 &+\cos(\theta/2)\sum_{k}{\cal G}_{k}(t)\ket{G}\ket{\boldsymbol{1}_k},
\end{aligned}
\label{eq:state}
\end{equation}
where $\ket{\boldsymbol{1}_k}$ is the cavity state with a single photon in mode $k$, i.e., $\ket{\boldsymbol{1}_k}=\hat{a}_k^{\dagger}\ket{\boldsymbol{0}}$ and ${{\cal G}_{k}}(t)$ is its probability amplitude.

The evolution of the state vector obeys the Schr\"{o}dinger equation. Substituting Eq. \eqref{eq:state} into the Schr\"{o}dinger equation $\left( i\dot{\ket{\psi}}=\hat{H}\ket{\psi}\right) $, we obtain the following  integro-differential equation for ${\cal E}(t)$:
\begin{equation}
\label{eq:Ediff}
\dot{\cal E}(t)+\cos^{4}(\eta /2)\int_{0}^{t}{d{{t}'}}F(t,{{t}'}){\cal E}({{t}'})=0
\end{equation}
where the kernel $F(t,{{t}'})$, is the correlation function defined in terms of continuous limits of the environment frequency as
\begin{equation}
\label{eq:kernel}
F(t,{{t}'})=\int_{0}^{\infty }{J({{\omega }_{k}}){{e}^{i({{\omega }_{D}}+{{\omega }_{L}}-{{\omega }_{k}})(t-{{t}'})}}}d{{\omega }_{k}},
\end{equation}
Here, $J({{\omega }_{k}})$ is the spectral density of reservoir modes. We choose a Lorentzian spectral density, which is typical of a structured cavity, whose form is
\begin{equation}
J({{\omega }_{k}})=\frac{1}{2\pi }\frac{\gamma {{\lambda }^{2}}}{[{{({{\omega }_{0}}-{{\omega }_{k}}-\delta )}^{2}}+{{\lambda }^{2}}]},
\end{equation}
where $\delta ={{\omega }_{0}}-{{\omega }_{c}}$ denotes the detuning between the center frequency of the cavity modes ${{\omega }_{c}}$  and ${{\omega }_{0}}$. The parameter $\gamma $ is related to the microscopic system-reservoir coupling constant, and $\lambda $ defines the spectral width of the coupling. It is noteworthy that the parameters $\gamma $and $\lambda $ are related to the reservoir correlation time ${{\tau }_{r}}$ and the qubit relaxation time ${{\tau }_{q}}$ as ${{\tau }_{r}}={{\lambda }^{-1}}$ and ${{\tau }_{q}}\approx {{\gamma }^{-1}}$ respectively \cite{Ali53}. Qubit-cavity weak coupling occurs for $\lambda >\gamma $ (${{\tau }_{r}}<{{\tau }_{q}}$); the opposite condition $\lambda <\gamma $ (${{\tau }_{r}}>{{\tau }_{q}}$) thus identifies strong coupling. The larger the cavity quality factor, the smaller the spectral width $\lambda $. 
 
 With such a spectral density, the kernel of Eq. (\ref{eq:kernel}) becomes 
\begin{equation}
F(t,{{t}'})=(\gamma \lambda /2){{e}^{-M(t-{t}')}}
\end{equation}
 with $M=\lambda -i({{\omega }_{D}}+\delta -\Delta )$. Substituting the resulting kernel into Eq. \eqref{eq:Ediff} yields
\begin{equation}
{\cal E}(t)=e^{-Mt/2}\left( \cosh ({\cal F}t/4)+\frac{2M}{{\cal F}}\sinh({\cal F}t/4)\right),
\end{equation}                          
 in which ${\cal F}=\sqrt{4{{M}^{2}}-2\gamma \lambda {{(1+\cos \eta )}^{2}}}$. The probability amplitude ${{\cal G}_{k}}(t)$ can be acquired as
\begin{equation}
\begin{aligned}
{{\cal G}_{k}}(t)&=-ig_k^*\cos^2(\eta/2)\Big( \Xi_+(t)+\Xi_-(t)\Big)
\end{aligned}
\end{equation} 
in which
\begin{equation}
\Xi_{\pm}(t)=\left(\frac{1}{2}\pm\frac{M}{{\cal F}} \right)\left( \frac{e^{(-\frac{M}{2}\pm\frac{{\cal F}}{4}-i(\delta_k+\Delta-\omega_D))t }-1}{-\frac{M}{2}\pm\frac{{\cal F}}{4}-i(\delta_k+\Delta-\omega_D)}\right)
\end{equation}
and ${{\delta }_{k}}={{\omega }_{k}}-{{\omega }_{0}}$. The time-dependent reduced density matrix of the qubit under the initial condition \eqref{eq:initialstate} can be obtained in the basis {$\left| E \right\rangle $, $\left| G \right\rangle $} as  
\begin{equation}
\hat{\rho}(t)=\begin{pmatrix}
  \cos^2(\theta/2)\left|{\cal E}(t)\right| ^2 & \frac{1}{2}\sin(\theta)e^{-i\phi}{\cal E}(t) \\
 \frac{1}{2}\sin(\theta)e^{i\phi}{\cal E}^*(t) & 1-\cos^2(\theta/2)\left|{\cal E}(t)\right| ^2
 \end{pmatrix}.
 \label{eq:red}
\end{equation}

\section{Spectrum}
\label{Spe}
By calculating ${{\cal G}_{k}}(t)$, one can obtain the spontaneous emission spectrum of the driven qubit as \cite{Ali54}: 
\begin{equation}
\label{spect}
S({{\delta }_{k}})=\frac{\gamma }{2\pi {{\left| {{g}_{k}} \right|}^{2}}}{{\left| {{\cal G}_{k}}(t\to \infty ) \right|}^{2}}
\end{equation}
Now, we present our numerical results based on expression (\ref{spect}) and show how the classical driving field can manipulate the spectrum. Fig. 2 illustrates the spontaneous emission spectrum $S({{\delta }_{k}})$ versus ${{\delta }_{k}}$, with $\Delta =0$, for different Rabi frequencies of the classical field, i.e., $\Omega$. Fig. 2 (a) shows that the spectrum gets an Autler$-$Townes doublet profile in the absence of the driving field. However, applying the classical field disturbs such a profile. As can be seen, increasing $\Omega $ weakens the left peak while not only strengthens the right peak but also shifts it farther away from ${{\delta }_{k}}=0$.

 \begin{figure}[h!]
   \centering
\includegraphics[width=0.6\textwidth]{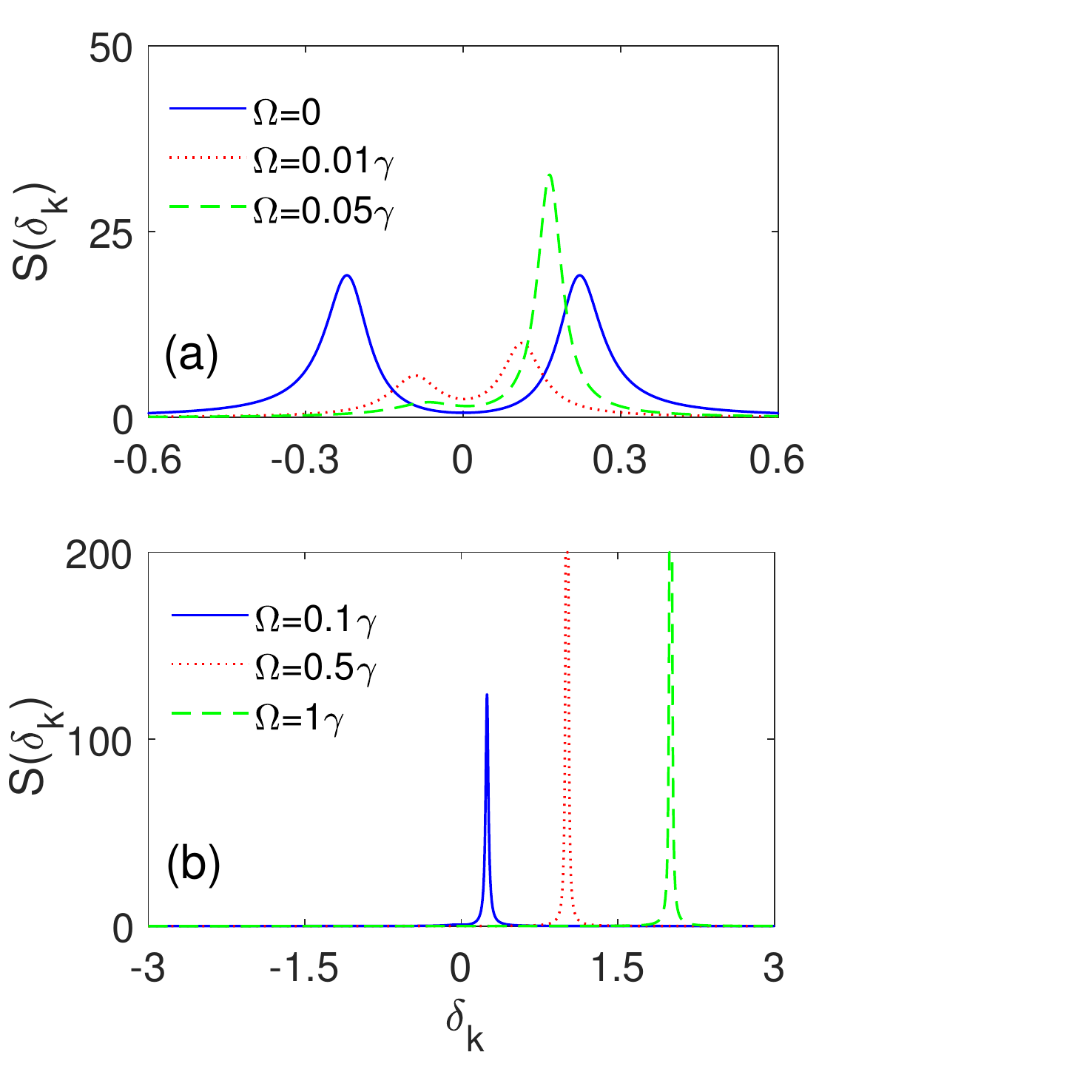}
   \caption{\label{Fig2} The spontaneous emission spectra $S(\delta_k)$ as a function of $\delta_k$ with $\Delta=0$ for different Rabi frequencies; (a) $\Omega=0\gamma$ (solid-blue line), $\Omega=0.01\gamma$ (dotted-red line), $\Omega=0.05\gamma$ (dashed-green line) and (b) $\Omega=0.1\gamma$ (solid-blue line), $\Omega=0.5\gamma$ (dotted-red line), $\Omega=1\gamma$ (dashed-green line). Others parameters are taken as: $\lambda=0.1\gamma$, $\delta=0$, $\theta=0$ and $\phi=0$.}
  \end{figure}

Fig. 3 displays the spontaneous emission spectrum $S({{\delta }_{k}})$ versus ${{\delta }_{k}}$ for different detunings between the qubit and the classical field. It is obvious; increasing $\Delta $ causes pattern of the Autler$-$Townes doublet fits to the spectrum profile. This result conveys the fact that increasing $\Delta $ weakens the interaction between the qubit and driving field. Hence, the profile for higher $\Delta $ has the most resemblance to the case in which the driving field is absent.

 \begin{figure}[h!]
   \centering
\includegraphics[width=0.6\textwidth]{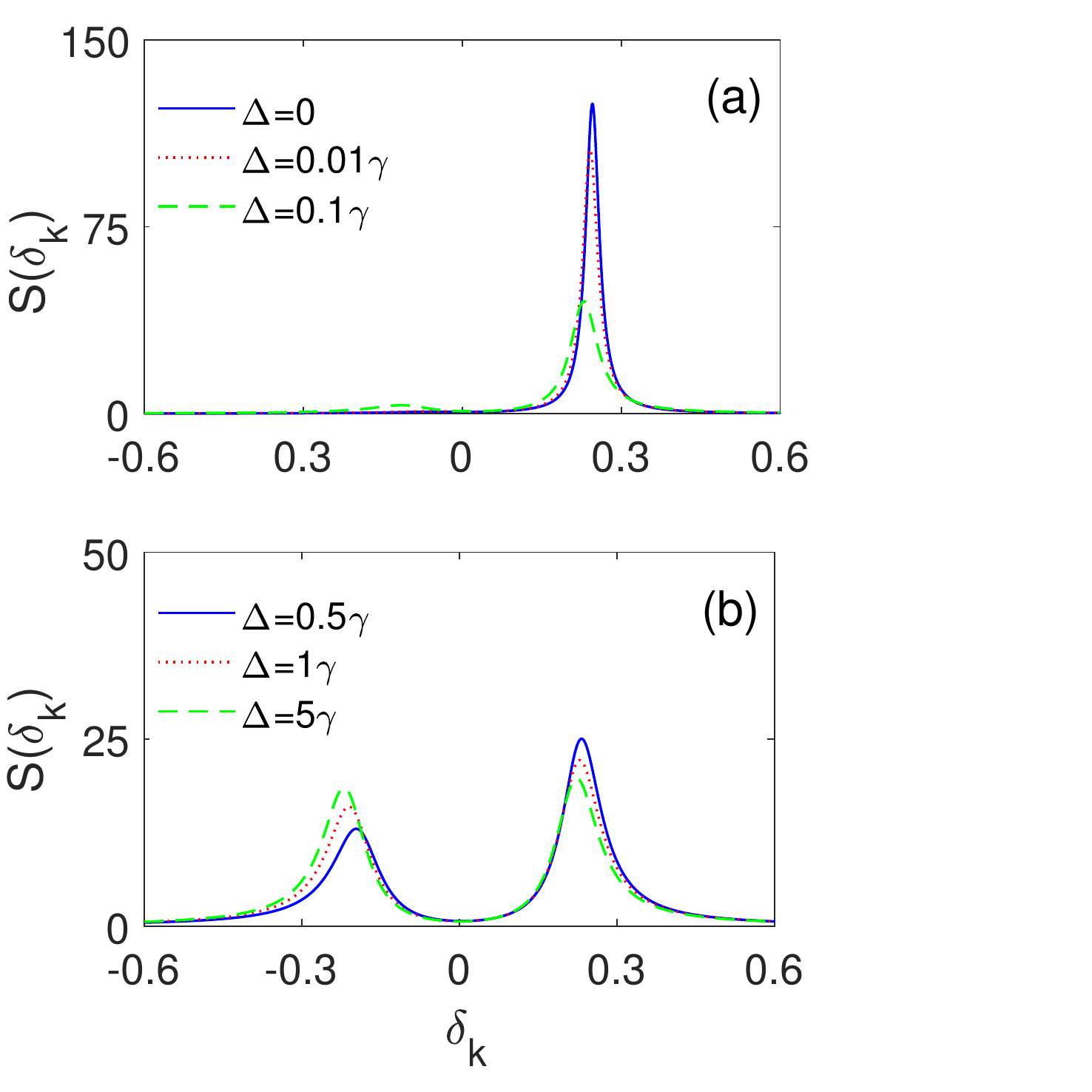}
   \caption{\label{Fig3} The spontaneous emission spectra $S(\delta_k)$ as a function of $\delta_k$ for different detunings between the qubit and the classical field; (a) $\Delta=0$ (solid-blue line), $\Delta=0.01\gamma$ (dotted-red line), $\Delta=0.1\gamma$ (dashed-green line) and (b) $\Delta=0.5\gamma$ (solid-blue line), $\Delta=1\gamma$ (dotted-red line), $\Delta=0.5\gamma$ (dashed-green line). Others parameters are taken as: $\lambda=0.1\gamma$, $\Omega=0.1\gamma$, $\delta=0$, $\theta=0$ and $\phi=0$.}
  \end{figure}

\section{Atom-Photon Entanglement}
\label{atmphy}
In this section, we discuss how the classical driving field can affect dynamics evolution of entanglement between the qubit and its radiative decay. It is worth recalling that the effect of coupling field on the dynamics of atom-photon entanglement in multi-level systems is studied \cite{Ali55,Ali56,Ali57,Ali58,Ali59}. For the bipartite systems such as atom-field systems, one can employ the von Neumann entropy of reduced density matrixes as a tool to quantify the entanglement between the subsystems. Generally, the von Neumann entropy $S$ of a system in quantum state $\rho $ is acquired by: 
\begin{equation}
S=-\text{Tr }\rho \ln \rho 
\end{equation}
It is crystal clear, this expression will be zero for any pure state and non-zero, $S\ne 0$, for mixed state. Araki and Lieb have demonstrated that for a bipartite quantum system composed of two subsystems A and F (say the atom and field) at any time t, the system and subsystems entropies satisfy the following triangle inequality condition \cite{Ali60}:
\begin{equation}
\left| {{S}_{A}}(t)-{{S}_{F}}(t) \right|\text{  }\le \text{  }{{S}_{AF}}(t)\text{  }\le \text{  }{{S}_{A}}(t)+{{S}_{F}}(t),\text{  }
\end{equation}
where ${{S}_{AF}}(t)=-\text{Tr }{{\rho }_{AF}}\ln {{\rho }_{AF}}$ denotes total entropy of the composite system and ${{S}_{A(F)}}(t)=-\text{T}{{\text{r}}_{A(F)}}\text{ }{{\rho }_{A(F)}}\ln {{\rho }_{A(F)}}$ are partial entropies corresponding to the reduced density matrix ${{\rho }_{A(F)}}=\text{T}{{\text{r}}_{F(A)}}{{\rho }_{AF}}$.

According to Eq. (17), if the field and the atom are initially prepared in pure states, the total entropy of the system becomes zero and remains constant. So partial entropies of both subsystems will evolve identically (${{S}_{A}}(t)={{S}_{F}}(t)$) after the interaction of two subsystems is switched on. Hence, entropy of the subsystems are recognized as an indication of the entanglement between subsystems. Namely, the higher the reduced quantum entropy, the greater the entanglement and vice versa \cite{Ali61,Ali62}.

In Fig. 4, we plot the entropy of the qubit versus the scaled time $\gamma t$, with $ \Delta =0 $, for different values of $ \Omega $. This figure discloses that, in the absence of the classical field, the qubit and its radiative decay become entangled instantly after the atom and cavity modes interact. However, entanglement damps with oscillatory behavior and eventually vanishes. On the other hand, applying the classical field and enhancing its intensity ($ \Omega $) not only causes the entanglement less fluctuate but it also remarkably prolongs the entanglement time.

The dynamic behavior of the entropy of the qubit for different values of $ \Delta $ is exhibited in Fig. 5. From this figure, one can figure out how increasing $\Delta $ leads to rapidly disappearing entanglement between qubit and its radiative decay. As mentioned above, the larger detuning $ \Delta $ makes the interaction between the qubit and the driving field weaker, thus the dynamics of entropy is mainly determined by the interaction between the qubit and reservoir. 

 \begin{figure}[h!]
   \centering
\includegraphics[width=0.6\textwidth]{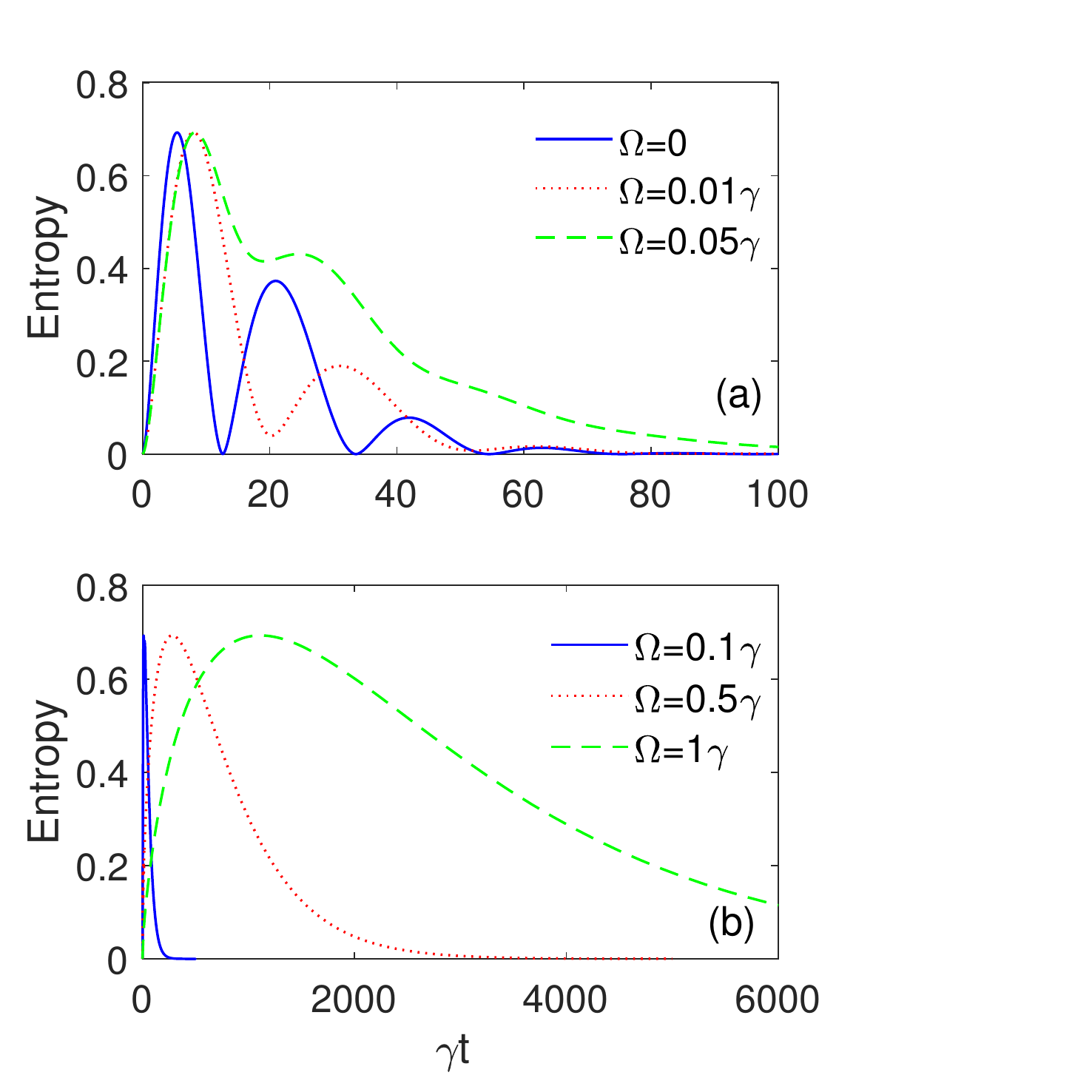}
   \caption{\label{Fig4} The entropy of the qubit as a function of scaled time $\gamma t$ for different Rabi frequencies; (a) $\Omega=0$ (solid-blue line), $\Omega=0.01\gamma$ (dotted-red line), $\Omega=0.05\gamma$ (dashed-green line) and (b) $\Omega=0.1\gamma$ (solid-blue line), $\Omega=0.5\gamma$ (dotted-red line), $\Omega=1\gamma$ (dashed-green line). Others parameters are taken as: $\lambda=0.1\gamma$, $\Delta=0$, $\delta=0$, $\theta=0$ and $\phi=0$.}
  \end{figure}
  
 \begin{figure}[h!]
    \centering
 \includegraphics[width=0.6\textwidth]{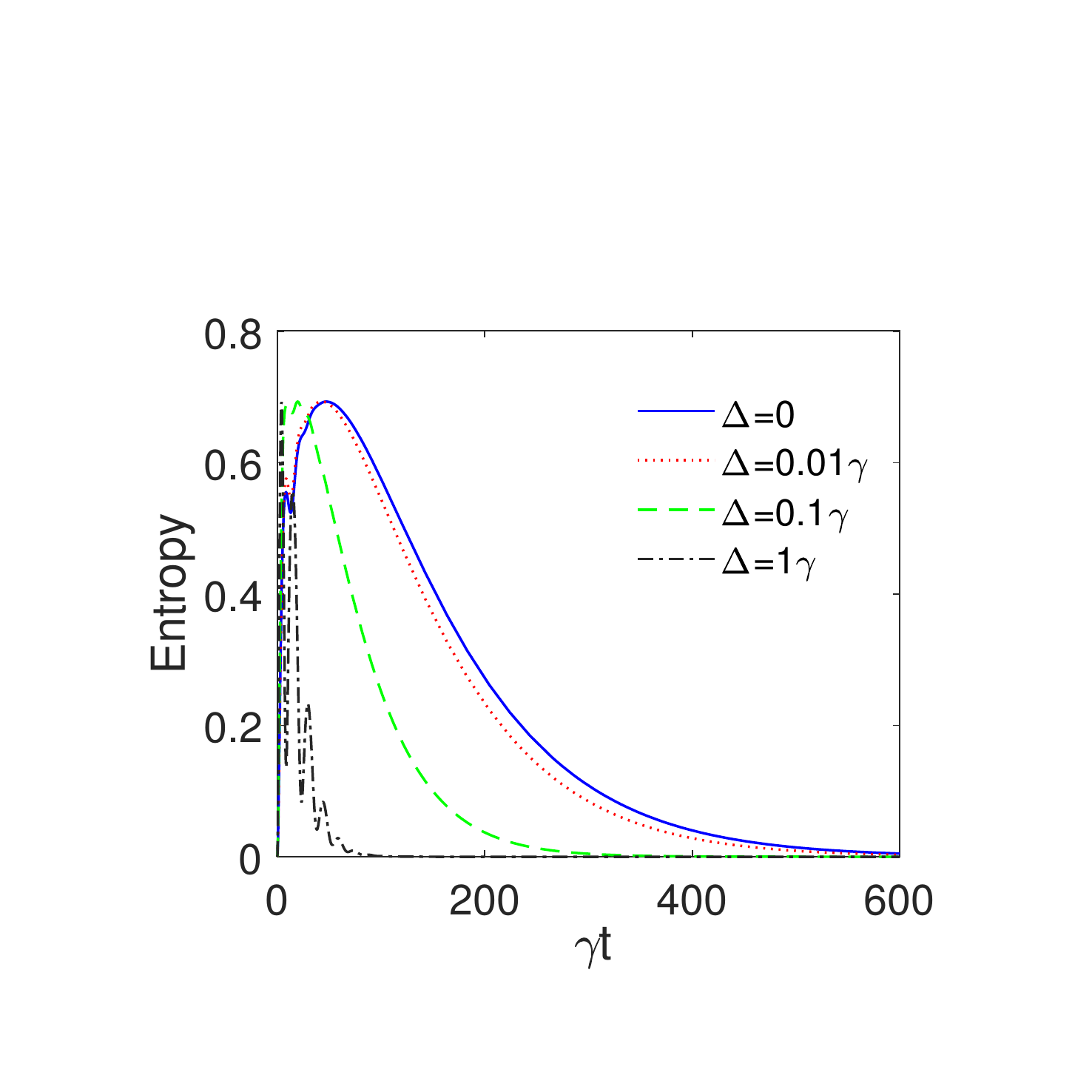}
    \caption{\label{Fig5} The entropy of the qubit as as a function of scaled time $\gamma t$ for different detunings between the qubit and the classical field; $\Delta=0$ (solid-blue line), $ \Delta=0.01\gamma $ (dotted-red line), $ \Delta=0.1\gamma $ (dashed-green line) and  $\Delta=1\gamma$ (dash-dotted-black line). Others parameters are taken as: $\lambda=0.1\gamma$, $\Omega=0.2\gamma$, $\theta=0$, $\phi=0$.}
   \end{figure} 
   
   \section{Entanglement Swapping}
   \label{EntSwa}
  According to the previous section, it is possible to create and control entanglement between the qubit and its radiative decay. Since creating and controlling entanglement between two qubits plays a crucial role in quantum information processing applications, a lot of methods have been proposed to realize it. Exchanging the entanglement stored between qubit-field in each cavity into qubit-qubit entanglement is regarded as a feasible solution. Hence, we address this issue by considering two identical qubit-field cavities (each one modelled in the section \ref{ModSol}). Note that, there is no direct interaction between the two qubit-field systems, therefore, their states are expected to remain separable:
            \begin{equation}\label{statet}
            \hat{\wp}(t)= \ket{\Psi(t)}\bra{\Psi(t)},
          \end{equation}
    in which $\ket{\Psi(t)}=\ket{\psi(t)}_1\otimes\ket{\psi(t)}_2$ where $\ket{\psi(t)}_i$ is the state vector of the $i$th qubit-field cavity reported in Eq. \eqref{eq:state}.
     \begin{figure}[h!]
       \centering
    \includegraphics[width=0.9\textwidth]{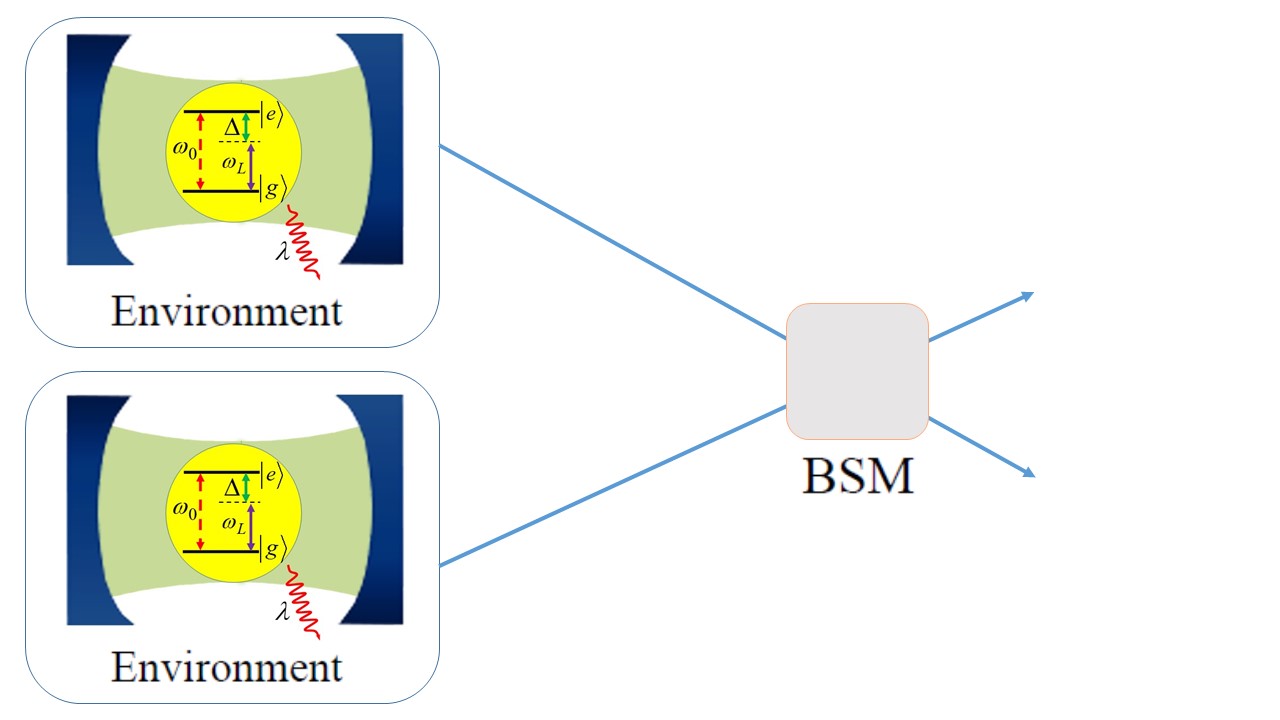}
       \caption{\label{Fig6} Pictorial representation of the entanglement swapping. Each qubit has been placed in its own cavity in the presence of dissipation. The BSM is performed on the photons leaving the environments which leads to the establishment of entanglement between the qubits.}
      \end{figure} 
   Then we perform a Bell state measurement on the field modes leaving the cavities (see Fig. \ref{Fig6}). Mathematically speaking, this is done by projection $\ket{\Psi(t)}$ onto one of the Bell states of the cavity fields. A glance at Eq. \eqref{eq:state} reveals that it is quite logical to consider the following Bell-like states of two-photon pairs \cite{Alireza15}
   \begin{subequations}
   \begin{eqnarray}
    \ket{\Psi^{\pm}}_\mathrm{F}&=&\frac{1}{\sqrt{2}}\left( \ket{\boldsymbol{0}}_{R_1}\ket{\boldsymbol{1}}_{R_2}\pm\ket{\boldsymbol{1}}_{R_1}\ket{\boldsymbol{0}}_{R_2} \right), \\
    \ket{\Phi^{\pm}}_\mathrm{F}&=&\frac{1}{\sqrt{2}}\left( \ket{\boldsymbol{0}}_{R_1}\ket{\boldsymbol{0}}_{R_2}\pm\ket{\boldsymbol{1}}_{R_1}\ket{\boldsymbol{1}}_{R_2} \right),
   \end{eqnarray}
   \end{subequations}
   in which $\ket{\boldsymbol{0}}_{R_i}$ is the vacuum state of the $i$th reservoir and 
   \begin{equation}
   \ket{\boldsymbol{1}}_{R_i}\equiv\sum_{k}\Theta_k\ket{\boldsymbol{1}_{k}}_i
   \end{equation}
   where $\sum_k |\Theta_k|^2 =1$ with $\Theta_k$ is related to the pulse shape associated with the incoming photons.
   
   It is possible to construct a desirable projection operator using the introduced Bell-type states, i.e.,  $P_\mathrm{F}=\ket{M}{}_\mathrm{FF}\bra{{}M}$ in which $M\in\left\lbrace \Psi^{\pm},\Phi^{\pm}\right\rbrace$. In this regard, by operating the projection operator onto $\ket{\Psi(t)}$, one can establish an entangled qubit-qubit state and leaves the filed states in a Bell-type state.

   In this paper, we choose the Bell state $\ket{\Psi^{-}}_\mathrm{F}$ to construct the following projection operator
   \begin{equation}
   P^-_\mathrm{F}=\ket{\Psi^-}{}_\mathrm{FF}\bra{{}\Psi^-},
   \end{equation}
   which its action on $\ket{\Psi(t)}$ in (\ref{statet}) leaves the field states in the Bell state $\ket{\Psi^-}{}_\mathrm{F}$ and establishes the following qubit-qubit state (after normalization):
   \begin{equation}\label{AAunnormPsi}
   \begin{aligned}
        \ket{\Psi_\mathrm{AA}(t)}&=P^-_F \ket{\Psi(t)}  \\
        &=\frac{1}{\sqrt{{\cal N}(t)}}\bigg\{  X(\theta_1,\theta_2,t)\Big(\ket{E}\ket{G}-\ket{G}\ket{E}\Big) \\
        &+  \Upsilon(\theta_1,\theta_2,\phi_1,\phi_2) \ket{G}\ket{G}\bigg\},
   \end{aligned}
   \end{equation}
   in which the normalization coefficient reads as
   \begin{equation}\label{NorCoeffPsi}
   \begin{aligned}
        {\cal N}(t)&= 2|X(\theta_1,\theta_2,t)|^2+|\Upsilon(\theta_1,\theta_2,\phi_1,\phi_2)|^2
   \end{aligned}
   \end{equation}
   where, we have defined
   \begin{subequations}
   \begin{eqnarray}
   X(\theta_1,\theta_2,t)&=&\cos(\theta_1/2)\cos(\theta_2/2){\cal E}(t), \label{eq:Theta} \\
   \Upsilon(\theta_1,\theta_2,\phi_1,\phi_2)&=&\sin(\theta_1/2)\cos(\theta_2/2)e^{i\phi_1} \nonumber \\  &-&\sin(\theta_2/2)\cos(\theta_1/2)e^{i\phi_2} \label{eq:Upsilon}
   \end{eqnarray}
   \end{subequations}
   In order to quantify the amount of entanglement between the two qubits, we use the concurrence which has been defined as \cite{Alireza13}
          \begin{equation}\label{con}
           E\left( \hat{\rho}(t)\right) =\mathrm{max}\{0,\sqrt{\lambda_1}-\sqrt{\lambda_2}-\sqrt{\lambda_3}-\sqrt{\lambda_4}\},
          \end{equation}
   where $\lambda_i$, $i=1,2,3,4$ are the eigenvalues (in decreasing order) of the Hermitian matrix
   $\hat{\rho}_{_\mathrm{AA}}\left(\sigma_1^y\otimes\sigma_2^y\hat{\rho}_{_\mathrm{AA}}^{*}\sigma_1^y\otimes\sigma_2^y\right)$ with $\hat{\rho}_{_\mathrm{AA}}^*$ the complex conjugate of $\hat{\rho}_{_\mathrm{AA}}$ and $\sigma_k^y:=i(\sigma_k-\sigma_k^\dag)$.  The concurrence varies between 0 (completely separable) and 1 (maximally entangled).
   For the state (\ref{AAunnormPsi}),  concurrence reads as
   \begin{equation}
   E\left( \hat{\rho}(t)\right)=\dfrac{2|X(\theta_1,\theta_2,t)|^2}{2|X(\theta_1,\theta_2,t)|^2+|\Upsilon(\theta_1,\theta_2,\phi_1,\phi_2)|^2}.
   \label{eq:conexPsi}
   \end{equation}
  
   The resulting concurrence is promising. As the Eq. (\ref{eq:conexPsi}) discloses, concurrence not only does not depend on the shape of the incoming photons(i.e., $\Theta_k$) but it also will be time-independent provided $|\Upsilon(\theta_1,\theta_2,\phi_1,\phi_2)|^2=0$, which means it remains always at its maximum value, i.e., 1. According to Eq. (\ref{eq:Upsilon}), it amounts to solve the following relation
   \begin{equation}
   \dfrac{1}{2}\Big( 1-\cos\theta_1\cos\theta_2 -\sin\theta_1\sin\theta_2\cos(\phi_1-\phi_2)\Big)=0.
   \end{equation}
   The above relation is fulfilled with the following set of solutions
   \begin{equation}
   \theta_1 =\theta_2 \ \text{and} \ \phi_1-\phi_2=2m\pi, \ \ m=0,\pm 1.
   \end{equation}
   This leads to the maximally entangled Bell state (up to an irrelevant global phase)
     \begin{eqnarray}\label{Bellstate1}
     \ket{\Psi^-}=\frac{1}{\sqrt{2}}(\ket{E}\ket{G}-\ket{G}\ket{E}).
     \end{eqnarray}

   As is seen, the concurrence (\ref{eq:conexPsi}) depends on the initial state of the two qubits. Therefore, one could obtain the relevant concurrence for specific initial states. However, thanks  to the parametrization \eqref{eq:initialstate}, it is possible to establish an input-independent (initial state-independent) dynamics of entanglement. This amounts to take a statistical average over the initial states: the concept of entangling power \cite{Alireza16}
   \begin{equation}
   {\mathfrak E}(t):=\int E\left(\rho(t)\right) \, d\mu( |\psi(0)\rangle),
   \label{eq:enpower}
   \end{equation}
   where $ d\mu( |\psi(0)\rangle)$ is the probability measure over the submanifold of product states in $\mathbb{C}^2\otimes \mathbb{C}^2$. The latter is induced by the Haar measure of ${\rm SU}(2) \otimes {\rm SU}(2)$. Specifically, referring to the parametrization of \eqref{eq:initialstate}, it reads
   \begin{equation}
   d\mu( |\psi(0)\rangle)=\frac{1}{16\pi^2}\prod\limits_{k=1}^2 \sin\theta_k\text{d}\theta_k\text{d}\phi_k.
   \end{equation}
   According to the above definition, the entangling power $\mathfrak E$ is normalized to 1. It is trivial that in this case it lies in $[0,1]$.\\
   
\begin{figure}[h!]
   \centering
   \includegraphics[width=0.6\textwidth]{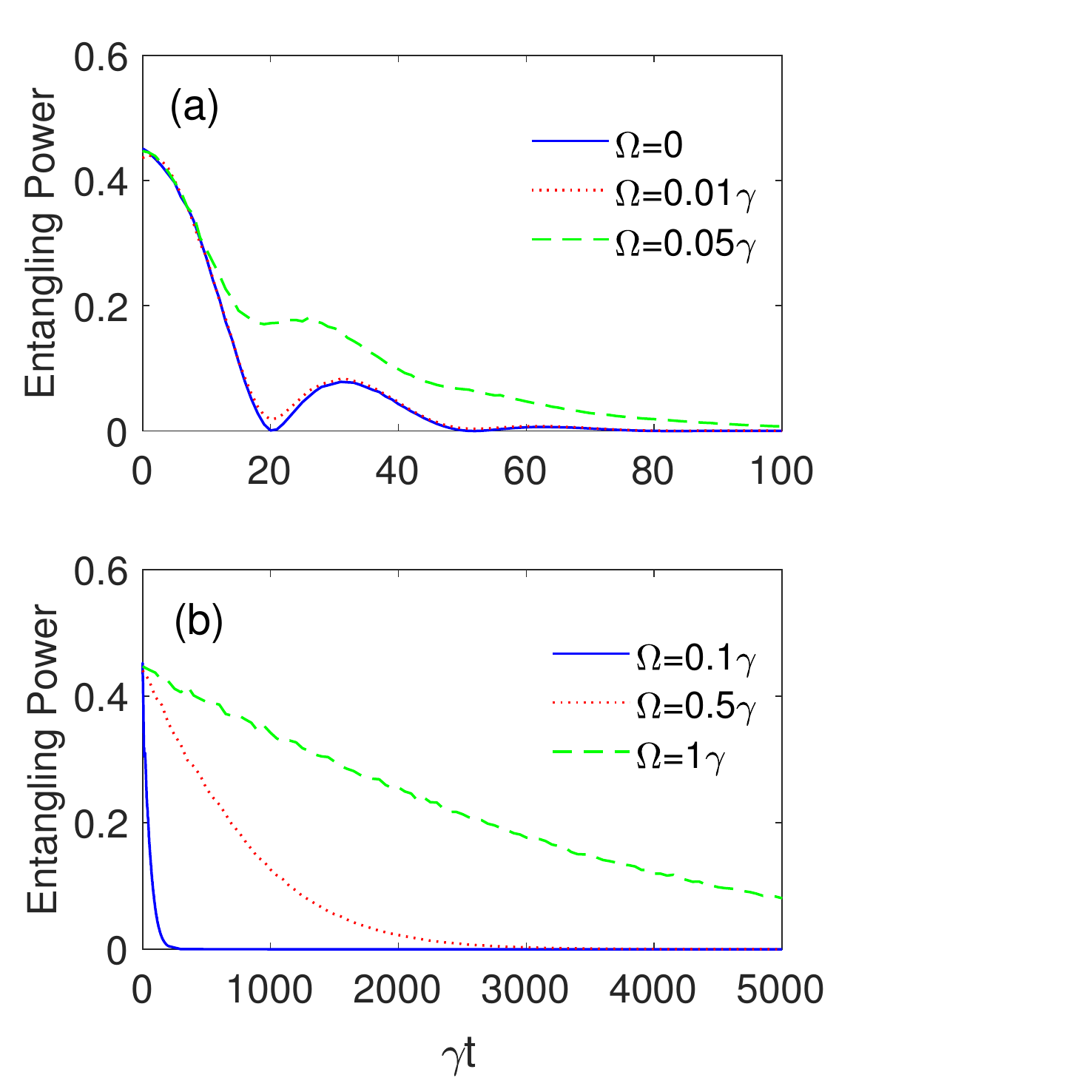}
      \caption{\label{Fig7} Time evolution of the entangling power of the qubit-qubit state after BSM ($P^-_\mathrm{F}=\ket{\Psi^-}{}_\mathrm{FF}\bra{{}\Psi^-}$) as a function of scaled time $\gamma t$ for different Rabi frequencies; (a) $\Omega=0$ (solid-blue line), $\Omega=0.01\gamma$ (dotted-red line), $\Omega=0.05\gamma$ (dashed-green line) and (b) $\Omega=0.1\gamma$ (solid-blue line), $\Omega=0.5\gamma$ (dotted-red line), $\Omega=1\gamma$ (dashed-green line). Others parameters are taken as: $\lambda=0.1\gamma$, $\Delta=0$ and $\delta=0$.}
     \end{figure}

   Fig. \ref{Fig7} illustrates the entangling power as a function of scaled time $\gamma t$ for different Rabi frequencies of the classical field, i.e., $\Omega$ with $\Delta=0$. It is evident that the Rabi frequency of the classical driving field has a beneficial effect on the survival of the entanglement. As can be seen, in the absence of the classical field the entangling power has a oscillation behaviour. In this case, the entanglement sudden death is clearly observed. However, switching the classical field on makes the entanglement survive at longer times. Actually, for enough large Rabi frequencies, it is possible to achieve a nearly stationary amount of the swapped entanglement between the two qubits, at least, at small times. These results are quite in consistent with the plots presented in Fig. \ref{Fig4}. 
   
\begin{figure}[h!]
       \centering
    \includegraphics[width=0.6\textwidth]{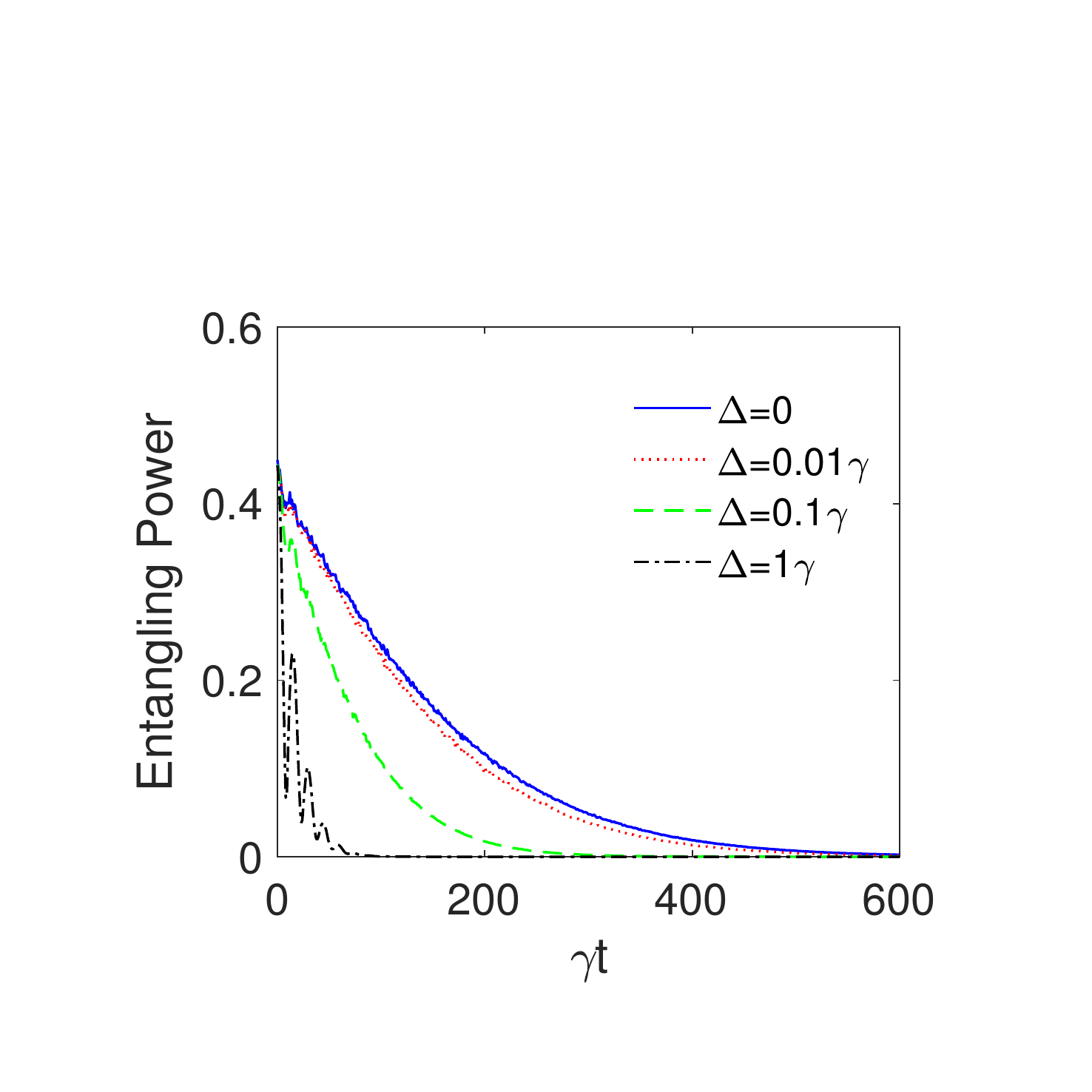}
       \caption{\label{Fig8} Time evolution of the entangling power of the atom-atom state after BSM ($P^-_\mathrm{F}=\ket{\Psi^-}{}_\mathrm{FF}\bra{{}\Psi^-}$) as a function of scaled time $\gamma t$ for different detunings between the qubit and the classical field;  $\Delta=0$ (solid-blue line), $\Delta=0.01\gamma$ (dotted-red line), $\Delta=0.1\gamma$ (dashed-green line) and  $\Delta=1\gamma$ (dash-dotted-black line). Others parameters are taken as: $\lambda=0.1\gamma$, $\Omega=0.2\gamma$ and $\delta=0$.}
      \end{figure}

   Eventually, we turn our attention to the non-resonant qubit-classical field interactions. In Fig. \ref{Fig8}, we display the effect of the detuning parameter $\Delta$ on the entangling power. On the contrary to the Rabi frequency, the detuning has a detrimental effect on the survival of the entangling power. For large values of $\Delta$, the oscillation of entanglement is clearly seen. Nonetheless, in this case, the entanglement sudden death does not happen. Again, the results are completely in consistent with those presented in Fig. \ref{Fig5}. 
     
\section{Conclusion}   
\label{Con}

In this paper, we have intended to study the effect of classical driving field on the spontaneous emission spectrum of the qubit as well as the entanglement dynamic between the qubit and its radiative decay in a dissipative cavity. We have found that an intense and resonant classical driving field not only remarkably modifies the spontaneous spectrum profile of the qubit but it also appreciably lengthens the time of entanglement between the qubit and its radiative decay. However, for the non-resonant interaction, increasing the detuning has detrimental effect.

We went further and investigated the possibility of the entanglement swapping between such two subsystems. To end this, we considered two similar cavities each consisting of a qubit interacting with a dissipative cavity under the action of a classical field. Then, we performed a Bell state measurement on the fields leaving cavities which establishes an entangled qubit-qubit state. We determined the situation in which a stationary long-lived maximally entangled qubit-qubit state can be generated. This occurs with the initial conditions $\theta_1=\theta_2$ and $\phi_1-\phi_2=2m\pi$ ($m=0,\pm 1$) which establishes the unique stationary state $ \ket{\Psi^-}=\frac{1}{\sqrt{2}}(\ket{E}\ket{G}-\ket{G}\ket{E})$. In order to study the dynamics of the swapped entanglement, we used the concurrence measure which depends on the initial  state of the two subsystems. Nonetheless, we established an input-independent dynamics by introducing the concept of the entangling power which takes a statistical average of the concurrence over the initial states. Our results suggest that the intensities of the classical driving fields have significant role on the survival of the swapped entanglement. It has been demonstrated that applying the classical fields and increasing their intensities will yield a long-lived swapped entanglement. However, increasing the detuning threatens survival of the entanglement.

At last, we should emphasize that our results could be employed in distribution of entangled states over long distances which has key role in quantum communications. This relies on the quantum repeater protocol which in turn has two bases: generating and swapping of the entanglement\cite{Alireza17}. Therefore, our results can be utilized to promote the efficiency of the quantum repeater protocol when the environmental effects cannot be neglected. This is left for future works.


%
%



\end{document}